\documentstyle[pre,preprint,aps]{revtex}

\tighten
\begin{document}
\draft
\preprint{adap-org/9505004}

\title{Hydrodynamic effects in density waves of granular flows}
\author{Kengo Ichiki
  \thanks{Electronic address: ichiki@cmpt01.phys.tohoku.ac.jp}
}
\address{Department of Physics, Tohoku University, Sendai 980-77, Japan}
\maketitle

\begin{abstract}
Granular flows in a narrow pipe are studied numerically
 by the model taking account of hydrodynamic effects
 of fluid surrounding particles.
In the simulations
density waves are observed over the wide range of the Stokes number,
 which represents the inertial effect of particles.
The mechanisms of formation of the density waves are considered and
 two types of density waves, which are observed in the systems with
 zero and finite Stokes numbers, are presented.
Power spectra of density waves obtained from the simulations
 with non-zero Stokes number
 show $1/f^\alpha$ power law
 which is similar to that in experiments.
\end{abstract}

\pacs{46.10.+z,47.55.Kf,05.40.+j}

\narrowtext

Granular materials are not only important in technology and industry,
 but also are very familiar to us.
We can see them in daily life, for example,
 sugar and salt in kitchen, sandy beach and soil on which we live.
Although these granular materials consist of simple components
 which are granular particles and surrounding fluid,
 they show a large variety of phenomena and
 attract a lot of interests of scientists
 \cite{jaeger1992,funtai1993,bideau,thornton,metha,hayakawa1995r}.
During the last decade, many people have studied a lot of subjects
 on granular flows such as
 vibrating beds\cite{evesque1989,taguchi1992,gallas1992},
 fluidized beds\cite{davidson,gidaspow,ichiki1995},
 hopper flows\cite{baxter1989,ristow1994},
 pipe flows \cite{schick1974,poschel1994,Lee,Peng,horikawa1995},
 and so on.

Pipe flow, which we discuss in this paper,
 seems to be the simplest system of them,
 in which granular particles is falling in a narrow pipe.
Even in such a simple system,
 it is observed in experiments that
 the uniform flow of particles is unstable and
 density waves appear \cite{schick1974,poschel1994,horikawa1995}.
In addition the fluctuation of density show $1/f^\alpha$ spectra
 \cite{schick1974,horikawa1995}.
By numerical simulations
 assuming mechanical dissipations
 such as roughness of walls and inelasticity of collisions,
 the creation of density waves \cite{poschel1994,Lee,Peng}
 and the appearance of $1/f^\alpha$ spectra \cite{Peng}
 were reported.
Studies of the clustering behavior of
 inelastic particles \cite{goldhirsch1993,mcnamara1994}
 are in the same context.

However one can expect naturally that
  the hydrodynamic effects of fluid surrounding particles
  must play an important role in the phenomena.
It is sufficient to remember a simple example
  that one particle is falling down by the gravity.
Situation is more difficult in the pipe flows
  because the hydrodynamic effects depend
  on the density of particle significantly \cite{hayakawa1995}.
Recently the present author and Hayakawa have constructed a model
  of granular particles
  that takes account of the hydrodynamic effects
  by applying the method of the Stokesian dynamics \cite{brady1988a},
  and have succeeded to create realistic bubbling and slugging flows
  in fluidized beds \cite{ichiki1995}.

The aim of this paper is
 to apply this model to the problems of pipe flows
 and to demonstrate the importance of the hydrodynamic effects.

Since the simulation method used in this paper is described in detail
 in Ref. \cite{ichiki1995},
 we briefly summarize the method.
Particles are driven by the gravity,
 hydrodynamic force from surrounding fluid
 and hard-core interaction in direct collisions between particles.
The equation of motion of particle $\alpha$ is given as
\begin{equation}
  St{d\over dt}{\bf v}^{(\alpha)}
  =
  {\bf F}_f^{(\alpha)}-{\bf e}_z,
  \label{eq:eq-of-motion-0}
\end{equation}
 where
 ${\bf v}^{(\alpha)}$ is the velocity of particle $\alpha$,
 ${\bf F}^{(\alpha)}_f$ is the hydrodynamic force,
 $-{\bf e}_z$ is the unit vector directed to the gravity
 and $St=mU_0/6\pi\mu a^2$ is the Stokes number.
In this paper we assume that all particles have same radius $a$ and
  the mass $m$.
In Eq. (\ref{eq:eq-of-motion-0}), quantities are nondimensionalized by
 the length $a$ and one-particle sedimentation velocity
 $U_0=m\tilde{g}/6\pi\mu a$,
 where $\tilde{g}$ is effective gravitational acceleration
 corrected by buoyancy
 and $\mu$ is the viscosity of the fluid.
Direct collisions between particles are considered to be elastic.
Particles interact with others through the hydrodynamic force
 which includes the many body effects of particle configuration.
In low Reynolds number limit,
 where the viscosity is dominant,
 ${\bf F}_f^{(\alpha)}$ can be written in the matrix form as
\begin{equation}
  {\bf F}^{(\alpha)}_f
  =
  -\sum_{\beta}{\sf R}^{(\alpha\beta)}\cdot{\bf v}^{(\beta)},
  \label{eq:eq-of-motion}
\end{equation}
 where ${\sf R}$ is so called resistance matrix scaled by $6\pi\mu a$.
As described in Ref. \cite{ichiki1995},
 the resistance matrix is constructed by the procedure of Stokesian dynamics
 \cite{brady1988a} with periodic boundary condition \cite{beenakker1986}.
Here we neglect the rotation of particles for simplicity.
In addition, we approximate (\ref{eq:eq-of-motion-0}) and
 (\ref{eq:eq-of-motion}) as follows,
\begin{eqnarray}
  St'{d\over dt}{\bf v}^{(\alpha)}
  &=&
  {\bf v}_t^{(\alpha)}-{\bf v}^{(\alpha)},
  \label{eq:eff-stokes}
  \\
  -{\bf e}_z
  &=&
  \sum_{\beta}
  {\sf R}^{(\alpha\beta)}\cdot{\bf v}_t^{(\beta)},
  \label{eq:term-vel}
\end{eqnarray}
 where ${\bf v}_t$ denotes the terminal velocity.
This approximation means that
 we neglect many body effects only in the relaxation process
 of particle velocity to the terminal velocity.
In this sense, $St'$ should be treated as the effective Stokes number.


In the simulation, we use narrow cells with periodic boundary condition.
For computational efficiency we only discuss the results
  on monolayer simulations
  where particles in each unit cell
  can move only on the plane parallel to the direction of gravity
  and the width of the cell is equal to the diameter of particles.
In fact there is no qualitative difference
  between the monolayer and full three-dimensional simulations.
All simulations presented in this paper are calculated as follows.
Particles are settled randomly with no initial velocities
  and the evolution obeys the equation of motion and the collision law.
In Table \ref{tab:settings} the parameters used in the simulations
  are summarized.

At first we investigate the effect of $St'$ which is the inertial effect
  of particles and the only parameter in our model.
Spatiotemporal patterns of density are shown in Fig. \ref{fig:pattern}
  (a),(b) and (c),
  where $St'$ is equal to $0$, $10$ and $100$ respectively.
We choose the other parameters as
  the number of particles in the unit cell $N=30$
  and the cell lengths $(L_x,L_y,L_z)=(6a,2a,80a)$.
The figure shows that
  uniform state is unstable in all case.
We can also observed that
  one sharp cluster are formed and
  the internal motions of particles in the cluster almost freeze in $St'=0$
  and otherwise broader cluster appears.
We note that small (large) $St'$ corresponds to
  the fluid with high (low) viscosity like water (air).
This means that in not only air but also water
  density waves can be observed.

Let us consider the mechanisms of dynamics of this behavior.
Our model consists of three types of mechanisms:
  inertial effects of particles,
  hydrodynamics interactions and hard-core collisions.
{}From the macroscopic point of view,
  the inertial effect and the collisions cause
  the advection and the diffusion of particles.
Hydrodynamic effect is realized as
  the mean velocity of particles depending on the local density,
  which is in general decreasing function of the density.
Therefore only the collisions
  can have the stabilizing effect on the disturbance of density.
We define the kinematic regime and the dynamic regime
  which are correspond to the cases
  where $St'$ is zero and finite respectively.
In the kinematic regime (Fig. \ref{fig:pattern} (a)),
  dynamical effects such as the advection and the collisions vanish
  and then systems are completely governed by (\ref{eq:term-vel}).
On the other hand in the dynamic regime (Fig. \ref{fig:pattern} (c)),
  the dynamical effects dominate.
The regime in Fig. \ref{fig:pattern} (b) may be the intermediate regime
  where both effects are important.

We vary parameter $N$ fixing $St'=10$ in order to examine
  the dependence of density wave on the particle density.
If the number of particles become large ($N=60$),
  internal motions of particles is weakened.
Thus dynamical effects are suppressed and
  particles are almost frozen in cluster like in the case of $St'=0$.
If the number of particles becomes small ($N=15$),
  which is (d) in Table \ref{tab:settings},
  density waves seem to be unstable though they are observed.

Let us examine the effect of wall confining fluid and particles.
For this purpose, we introduce the fixed particles
  which form vertical wall.
It is found from the simulation that
  the wall enhances the horizontal motions and the collisional effects
  when $St'$ is finite.
The system with wall behaves like that without wall but
  with larger $St'$.

Next we investigate the power spectra obtained from the simulations
  in Table \ref{tab:settings}
  (Fig. \ref{fig:power-spectrum}).
These spectra are calculated as follows.
At first we divide horizontally the unit cell into 20 sub cells
  and calculate the time series of density in each sub cell.
Then we calculate the power spectra of each sub cell
  by a fast Fourier transform with the Parzen window
  for the last $16384$ time series of the simulation
  and average them over 20 sub cells.
Frequencies are scaled by $f_0$ which is the lowest frequency.
The simulation (e) is presented as the example with no density waves
  for comparison,
  which correspond to the case of $St'=\infty$ and
  is performed without the hydrodynamic effects and the gravity
  but with the initial velocity as their terminal velocities
  in the case of $St'=10$.
We can see that power laws are observed for the simulations
 with finite $St'$
 in the range between $f_c=V_c/L_z$ and $f_s=\bar{V}/4a$,
 where $V_c$ and $\bar{V}$ are
 the mean velocity of the cluster and the particles respectively,
 $L_z$ is system size in direction of the gravity
 and $4a$ is the length of the sub cells
 used in the calculation of spectra.
Therefore $f_c$ and $f_s$ correspond to
 the return time of cluster and the characteristic time in small scale motion
 respectively.
The exponents in the range are shown in Table \ref{tab:results}.

This kind of spectra of density fluctuation have been observed
  in hourglass twenty years ago \cite{schick1974}.
In a recent experiment of pipe flow \cite{horikawa1995},
 similar power spectra have been obtained and
 the high-frequency limit in them may corresponds to $f_s$
 in this simulation.
{}From the meaning of $f_c$ and $f_s$,
 it is suggested that
 the power law reflects the behavior in relatively small scale
 rather than the cluster itself.
The existence of low-frequency limit in Fig. \ref{fig:power-spectrum}
 may be due to the limit of system size
 or the periodicity in the simulations.
Therefore we need to construct the model without periodic boundary condition
 in order to investigate the behavior of larger scale like clusters
 and the relation to the real systems.


In conclusion, by means of the numerical simulations \cite{ichiki1995},
 it is found that in pipe flows the hydrodynamic effects of the fluid
 surrounding particles play an important role
 and density waves appear in the wide range of $St$.
This means that in the water as well as in the air
 density waves can be observed.
Two mechanisms to form density waves,
 which are inertial effects of particles and the hydrodynamic interactions,
 are suggested.
In the case of finite $St'$,
 $1/f^\alpha$ power laws, which may represent the behavior of particles
 falling from the cluster, are observed.


Finally we note that
 there are a lot of resemblances between the pipe flows
 and the traffic flows in an expressway.
The kinematic waves have been studied as the model of traffic flows
 \cite{Whitham}.
In addition our model is formally similar to
 the model of traffic flows presented in Ref. \cite{Bando}
 and $1/f$ power spectra are also observed in the real traffic
 \cite{Musha}
 and in the numerical model \cite{nagel1995}.

The author thanks
 Hisao Hayakawa and Toshio Tsuzuki for stimulating discussions,
 Yoshi-hiro Taguchi, Akio Nakahara and Mitsugu Matsushita
 for helpful comments and encouragements
 and Teruhisa S. Komatsu for useful suggestion on the resemblance to
 traffic flow.
Most of simulations in this paper have been done using the facilities of
the Supercomputer Center, Institute of Solid State of Physics, University
of Tokyo.


\begin{table}
  \caption{
    The parameters of simulations.
    $St'$ is the effective Stokes number in Eq. (\protect\ref{eq:eff-stokes}),
    $N$ is the number of particles in the unit cell,
    $L_x$,$L_y$ and $L_z$ are the lengths of the unit cell.
    }
  \label{tab:settings}
  \begin{tabular}{lccccl}
    & $St'$ & $N$ & $L_x$ & $L_y$ & $L_z$ \\
    \tableline
    (a) & $0.0$ & $30$ & $6$ & $2$ & $80$ \\
    (b) & $10.0$ & $30$ & $6$ & $2$ & $80$ \\
    (c) & $100.0$ & $30$ & $6$ & $2$ & $80$ \\
    (d) & $10.0$ & $15$ & $6$ & $2$ & $80$ \\
    (e) & $\infty$ & $30$ & $6$ & $2$ & $80$
  \end{tabular}
\end{table}

\begin{table}
  \caption{
    The characteristic velocities and frequencies
    and the exponents of power laws.
    $V_c$ and $\bar{V}$ is scaled by one-particle sedimentation velocity
    $U_0$,
    and $f_c$ and $f_s$ is scaled by the lowest frequency $f_0$.
    }
  \label{tab:results}
  \begin{tabular}{lccccc}
    & $V_c$ & $\bar{V}$ &  $f_c$ & $f_s$ & $\alpha$\\
    \tableline
    (b) & $0.12$ & $0.28$ & $8$ & $350$ & $-1.47\pm 3\times 10^{-2}$ \\
    (c) & $0.21$ & $0.31$ & $13$ & $380$ & $-1.40\pm 3\times 10^{-2}$ \\
    (d) & $0.18$ & $0.43$ & $11$ & $510$ & $-1.32\pm 3\times 10^{-2}$
  \end{tabular}
\end{table}

\begin{figure}
  \caption{
    Spatiotemporal patterns of density
    for the first 6000 scaled time
    with (a) $St'=0$,(b) $St'=10$ and (c) $St'=100$.
    Other parameters are listed in Table \protect\ref{tab:settings}.
    Dark regions correspond to high densities.
    }
  \label{fig:pattern}
\end{figure}

\begin{figure}
  \caption{
    Power spectra $S(f)$ of density fluctuation of the simulations in
    Table \protect\ref{tab:settings}.
    These spectra are calculated by a fast Fourier transform with
    the Parzen window.
    Frequency is scaled by the lowest one.
    Arrows correspond to $f_c$ and $f_s$ in Table \protect\ref{tab:results}.
    }
  \label{fig:power-spectrum}
\end{figure}

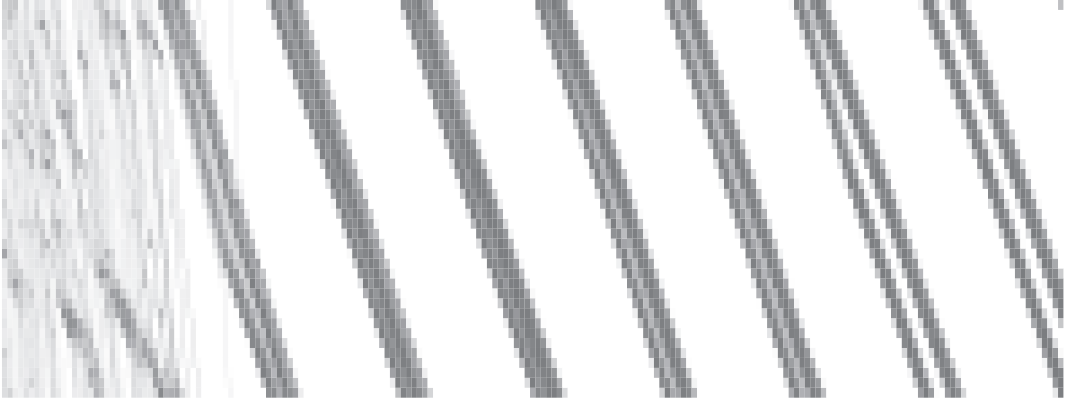
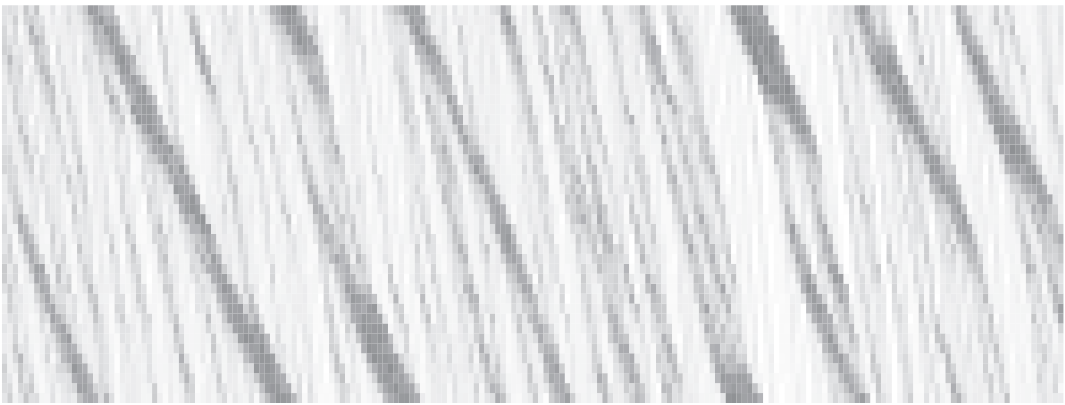
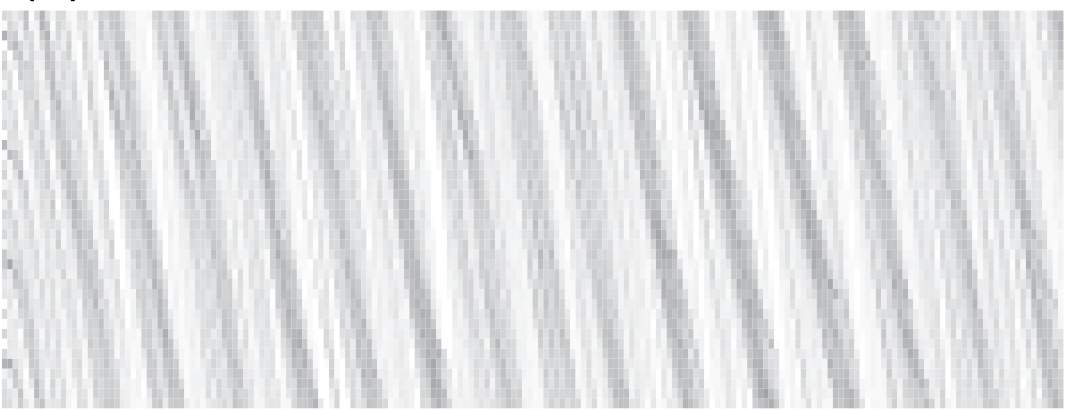
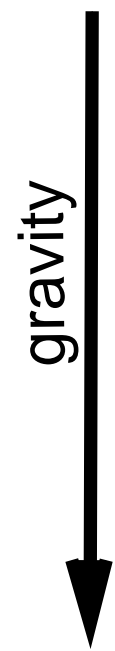
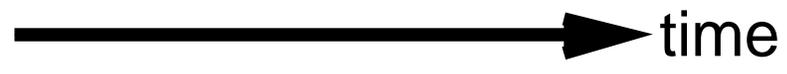
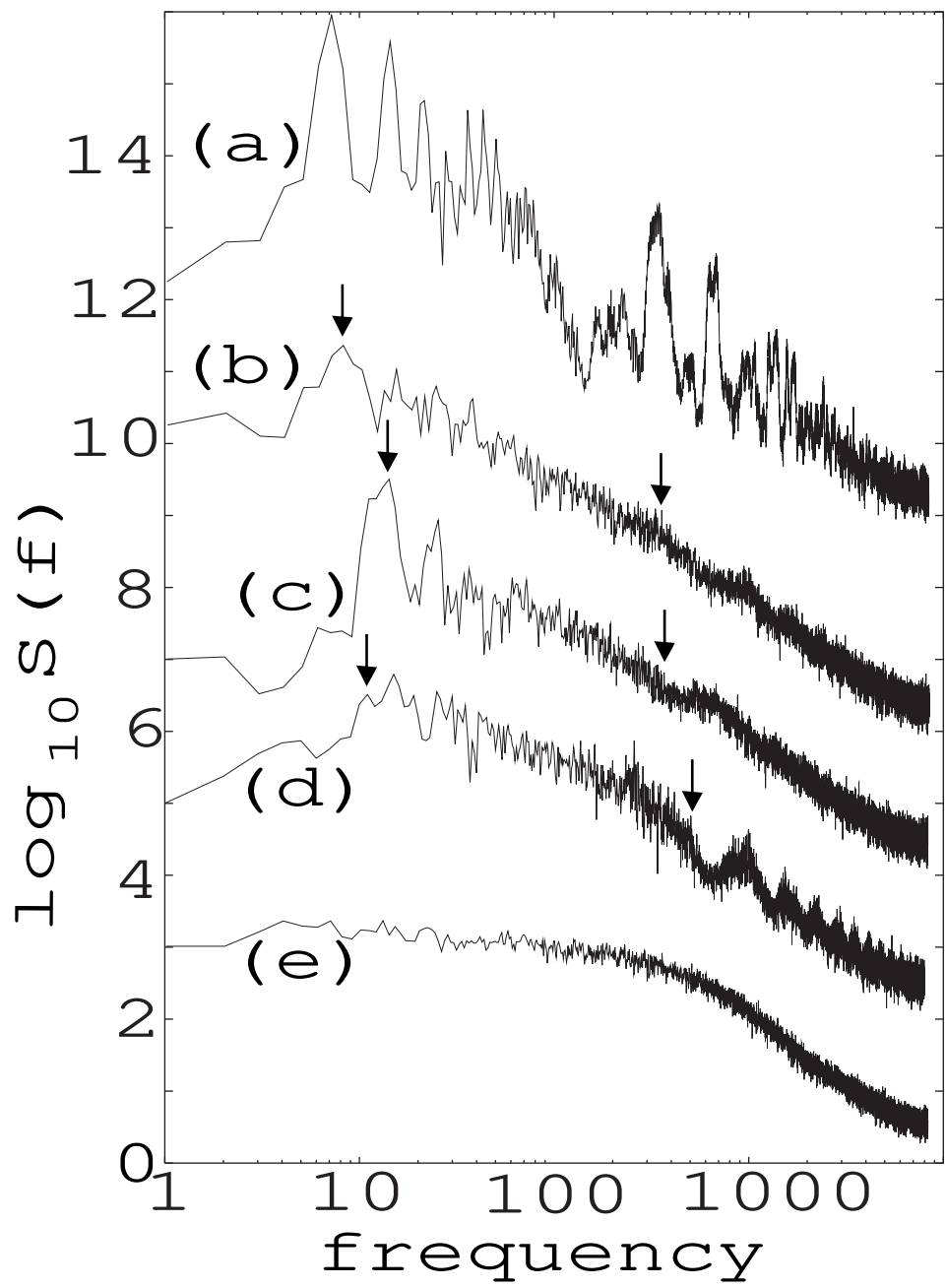
\end{document}